\documentclass[12pt,preprint]{aastex}

\newcommand{\hst}{{\it HST}}

\newcommand{\mps}{m~s$^{-1}$}
\newcommand{\kps}{km~s$^{-1}$}
\newcommand{\Mjup}{\mbox{$M_{\rm J}$}}
\newcommand{\Rjup}{\mbox{$R_{\rm J}$}}
\newcommand{\Mplan}{\mbox{$M_{\rm p}$}}

\newcommand{\Mstar}{\mbox{$M_{\rm *}$}}
\newcommand{\Rstar}{\mbox{$R_{\rm *}$}}
\newcommand{\Pstar}{\mbox{$P_{\rm *}$}}
\newcommand{\Msun}{\mbox{$M_\odot$}}
\newcommand{\Rsun}{\mbox{$R_\odot$}}
\newcommand{\Teff}{\mbox{$T_{\rm eff}$}}
\newcommand{\Logg}{\mbox{$\log g$}}

\newcommand{\MonH}{\mbox{[M/H]}}
\newcommand{\NaH}{\mbox{[Na/H]}}
\newcommand{\SiH}{\mbox{[Si/H]}}
\newcommand{\TiH}{\mbox{[Ti/H]}}
\newcommand{\FeH}{\mbox{[Fe/H]}}
\newcommand{\NiH}{\mbox{[Ni/H]}}
\newcommand{\Vsini}{\mbox{$v\sin i$}}
\newcommand{\kms}{\mbox{km s$^{-1}$}}
\newcommand{\rchisq}{\mbox{$\chi_{\rm r}^2$}}
\newcommand{\Mgb}{\mbox{\ion{Mg}{1} b}}
\newcommand{\Ytwo}{Y$^2$}

\newcommand{\xonb}{XO-4b}
\newcommand{\xon}{XO-4}
\newcommand{\vMs}{1.32}		
\newcommand{\eMs}{0.02}
\newcommand{\vRs}{1.55}		
\newcommand{\eRs}{0.05}
\newcommand{\vaoRs}{7.7}		
\newcommand{\eaoRs}{0.2}
\newcommand{\vRpoRs}{0.089}		
\newcommand{\eRpoRs}{0.001}
\newcommand{\vAge}{2.1}
\newcommand{\eAge}{0.6}
\newcommand{\sptype}{F5V}
\newcommand{\vTeff}{5700}		
\newcommand{\eTeff}{70}		
\newcommand{\vfeoh}{-0.04}		
\newcommand{\efeoh}{0.03}		
\newcommand{\vlogg}{4.18}		
\newcommand{\elogg}{0.07}		
\newcommand{\vVsini}{8.8}		
\newcommand{\eVsini}{0.5}		
\newcommand{\vrvK}{163}		
\newcommand{\ervK}{16}
\newcommand{\vDs}{293}		
\newcommand{\eDs}{19}
\newcommand{\vjd}{2454485.9322}	
\newcommand{\ejd}{0.0004}	
\newcommand{\vap}{0.0555}	
\newcommand{\eap}{0.0011}
\newcommand{\vbp}{0.18}	
\newcommand{\ebp}{0.14}
\newcommand{\vperiod}{4.12502}	
\newcommand{\eperiod}{0.00002}
\newcommand{\vMp}{1.72}		
\newcommand{\eMp}{0.20}	
\newcommand{\vRp}{1.34}		
\newcommand{\eRp}{0.048}
\newcommand{\vincl}{88.7}		
\newcommand{\eincl}{1.1}

\newcommand{\gsc}{GSC 03793-01994}

\slugcomment{Submitted for publication in the Astrophysical Journal}

\received{2008 May 01}
\begin{document}

\title{XO-4b: An Extrasolar Planet Transiting an F5V Star}

\author{
P.~R.~McCullough\altaffilmark{1},
Christopher~J.~Burke\altaffilmark{1},
Jeff~A.~Valenti\altaffilmark{1},
Doug~Long\altaffilmark{1},
Christopher~M.~Johns-Krull\altaffilmark{2,3},
P.~Machalek\altaffilmark{1,3,4},
K.~A.~Janes\altaffilmark{5}, 
B.~Taylor\altaffilmark{6},
J.~Gregorio\altaffilmark{7},
C.~N.~Foote\altaffilmark{8},
Bruce~L.~Gary\altaffilmark{9},
M.~Fleenor\altaffilmark{10},
Enrique~Garc\'{i}a-Melendo\altaffilmark{11},
T.~Vanmunster\altaffilmark{12}
}

\email{pmcc@stsci.edu}

\altaffiltext{1}{Space Telescope Science Institute, 3700 San Martin Dr., Baltimore MD 21218}
\altaffiltext{2}{Dept. of Physics and Astronomy, Rice University, 6100 Main Street, MS-108, Houston, TX 77005}
\altaffiltext{3}{Visiting Astronomer, McDonald Observatory, which is operated by the University of Texas at Austin.}
\altaffiltext{4}{Department of Physics and Astronomy, Johns Hopkins University, 3400 North Charles Street, Baltimore, MD 21218}
\altaffiltext{5}{Boston University, Astronomy Dept., 725 Commonwealth Ave., Boston, MA 02215}
\altaffiltext{6}{Lowell Observatory, 1400 W. Mars Hill Rd., Flagstaff, AZ 86001}
\altaffiltext{7}{Obs. Atalaia, Alcabideche, Portugal}
\altaffiltext{8}{Vermillion Cliffs Observatory, Kanab, UT}
\altaffiltext{9}{Hereford Arizona Observatory, 5320 E. Calle Manzana, Hereford, AZ 85615}
\altaffiltext{10}{Volunteer Observatory, Knoxville, TN}
\altaffiltext{11}{Esteve Duran Observatory Foundation, Montseny 46, 08553 Seva, Spain}
\altaffiltext{12}{CBA Belgium Observatory, Walhostraat 1A, B-3401 Landen, Belgium}

\begin{abstract}
We report the discovery of the planet \xonb, which transits the star \xon\ (\gsc, V=10.7, \sptype).
Transits are 1.0\% deep and 4.4 hours in duration.
The star \xon\ has
  a mass of $\vMs\pm\eMs$ \Msun, 
  a radius of $\vRs\pm\eRs$ \Rsun,
  an age of $\vAge\pm\eAge$ Gyr,
  a distance of $\vDs\pm\eDs$ pc,
  an effective temperature of $\vTeff\pm\eTeff$ K,
  a logarithmic iron abundance of $\vfeoh\pm\efeoh$ relative to solar,
  a logarithmic surface gravity in cgs units of $\vlogg\pm\elogg$,
  and
  a projected rotational velocity of $\vVsini\pm\eVsini$ \kps.
The star \xon\ has periodic radial velocity variations with a semi-amplitude of $\vrvK\pm\ervK$ \mps, due to the planet \xonb.
The planet \xonb\ has
  a mass of $\vMp\pm\eMp$ \Mjup,
  a radius of $\vRp\pm\eRp$ \Rjup,
  an orbital period of $\vperiod\pm\eperiod$ days,
  and
  a heliocentric Julian date at mid-transit of $\vjd\pm\ejd$.
We analyze scintillation-limited differential R-band photometry of \xonb\ in
transit made with a 1.8-m telescope under photometric conditions,
yielding photometric precision of 0.6 to 2.0 millimag per one-minute interval.
The declination of \xon\ places it within the continuous viewing zone of
the Hubble Space Telescope (\hst), which permits observation without
interruption caused by occultation by the Earth.
Because the stellar rotation periods of the three hottest stars
orbited by transiting gas-giant planets are 2.0, 1.1, and 2.0 times
the planetary orbital periods, we note the possibility of resonant
interaction.
\end{abstract}


\section{Introduction}

During 2007, the number of reported planets that transit stars brighter than
$V=13$ increased markedly from 9 to 23
planets\footnote{Extrasolar Planets Encyclopaedia, http://exoplanet.eu}.
The number of transiting planets reported in the first half of 2008 has
already exceeded the total number reported in 2007.
This rapid increase in discovery rate is due in part to recently matured search
techniques and analysis software applied to thousands of observations
of hundreds of thousands of stars by multiple research groups.
The planet \xonb\ reported here is another in a sequence of discoveries
by the XO project (McCullough et al.\ 2005), facilitated by a collaboration
between professional and amateur astronomers.\footnote{This paper
includes data taken at the Haleakala summit (maintained by the University
of Hawaii), Lowell Observatory, McDonald Observatory (operated by the
University of Texas at Austin), and numerous backyard observatories.}

The star \xon\ is bright enough to appear in a few all-sky surveys with
the following identifiers:
  \gsc\ (Morrison et al.\ 2001),
  2MASS J07213317+5816051 (Cutri et al.\ 2003),
  and
  TYC 3793--01994--1 (H{\o}g et al.\ 2000).
For \xon\ the Tycho Catalog reports
  right ascension
    $\alpha = 7^{\rm h}21^{\rm m}33^{\rm s}.159\pm0.026\arcsec$,
  declination
    $\delta = +58\arcdeg16\arcmin4\farcs98 \pm 0.028\arcsec$,
  and proper motions 
    $\mu_\alpha = -18.1 \pm 2.9$ mas yr$^{-1}$
  and
    $\mu_\delta = -4.0 \pm 2.9$ mas yr$^{-1}$,
all for equinox and epoch J2000.

Several recently discovered transiting planets expanded the parameter space
of \textit{transiting} planets in one way or another at the time of their discovery.
For example, GJ 426b (Gillon et al.\ 2007) is the smallest and least massive
transiting planet, and also the only one known to transit an M dwarf.
XO-3b (Johns--Krull et al.\ 2008) is the most massive ($\Mplan=12\Mjup$).
HAT--P--2b (Bakos et al.\ 2007) had the largest eccentricity ($e=0.52$), until
HD 17156b (Barbieri et al.\ 2007) was discovered
with a larger eccentricity ($e=0.67$) and the longest known period
($P=21$ days), demonstrating the ephemeral nature of such records.
As Harwit (1981) discusses, the limits of the extent of any parameter space
begin to show themselves not first in the outliers themselves, but in the
``repetitions'' of discoveries. Simply put, if nearly every new discovery is
unique in some significant way, then clearly we have not yet begun to bound
the parameter space.
\xonb\ is not particularly unusual with respect to other known transiting
planets, except perhaps that it orbits a relatively warm star. Only
HAT--P--6b (Noyes et al.\ 2008) and XO-3b orbit slightly warmer stars.
In fact, the HAT--P--6 and \xon\ systems are quite similar.

To stimulate additional observations and analysis of the \xon\ system,
we report here the characteristics of both the star and the planet.
In Section \ref{sec:obs} we describe the observations, which include time-series
photometry, out-of-transit
${\rm BVR_CI_C}$ photometry, and spectroscopy with and without
iodine absorption lines superposed.
In Section \ref{sec:analysis} we analyze the data and derive the ephemeris
of the transits, measure time-dependent radial velocities, and the
physical characteristics of the star \xon, using both its spectrum
and its light curve. 
Section \ref{sec:lightcurve}
applies the numerical models commonly used for transits
to scintillation-limited photometry.
Section \ref{sec:discussion} discusses the results and summarizes the conclusions.

\section{Observations\label{sec:obs}}

\subsection{Survey and Extended-Team Photometry}

\begin{figure}
\plotone{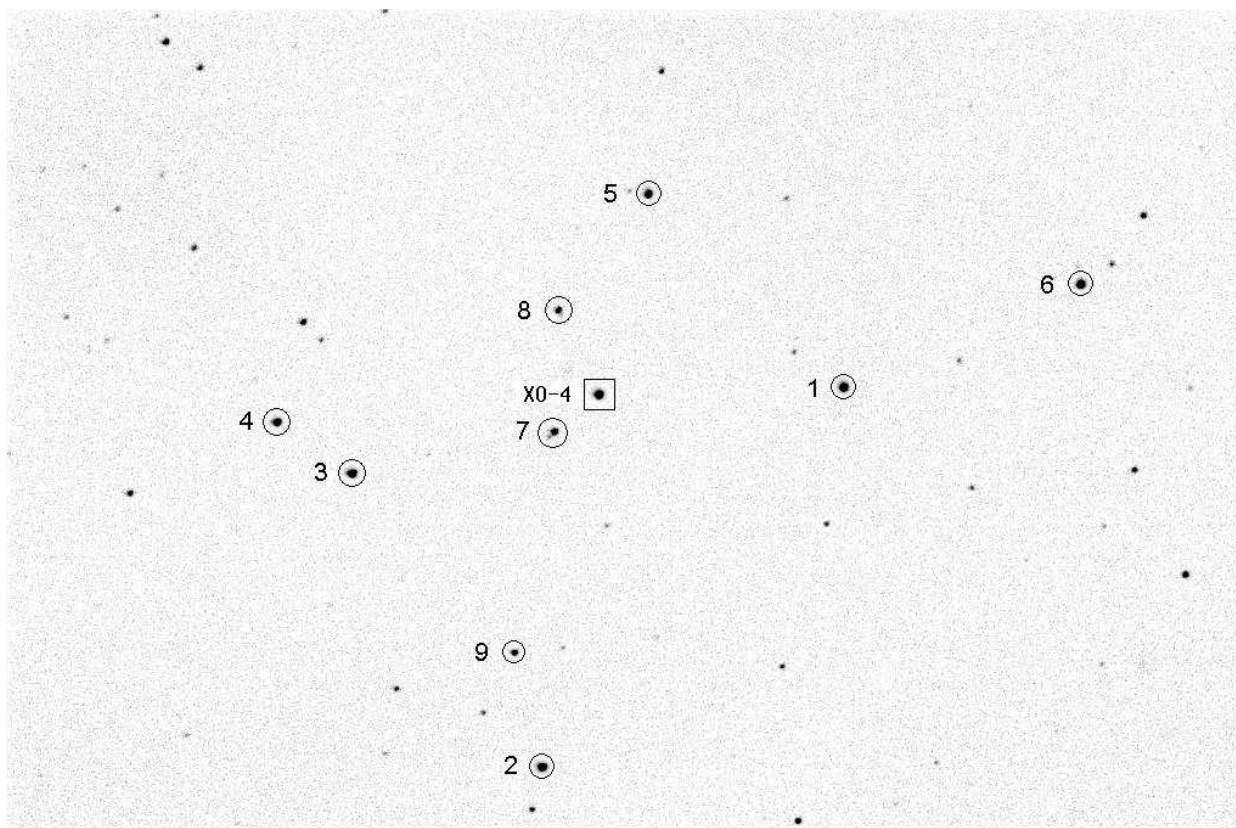}
\caption{
\xon\ is inside the square.
Stars from Table \ref{table:allsky} are circled and numbered. 
The CCD image is 16\arcmin\ by 11\arcmin\ with North up and East to the left. 
\label{fig:finder}}
\end{figure}

\begin{figure}
\plotone{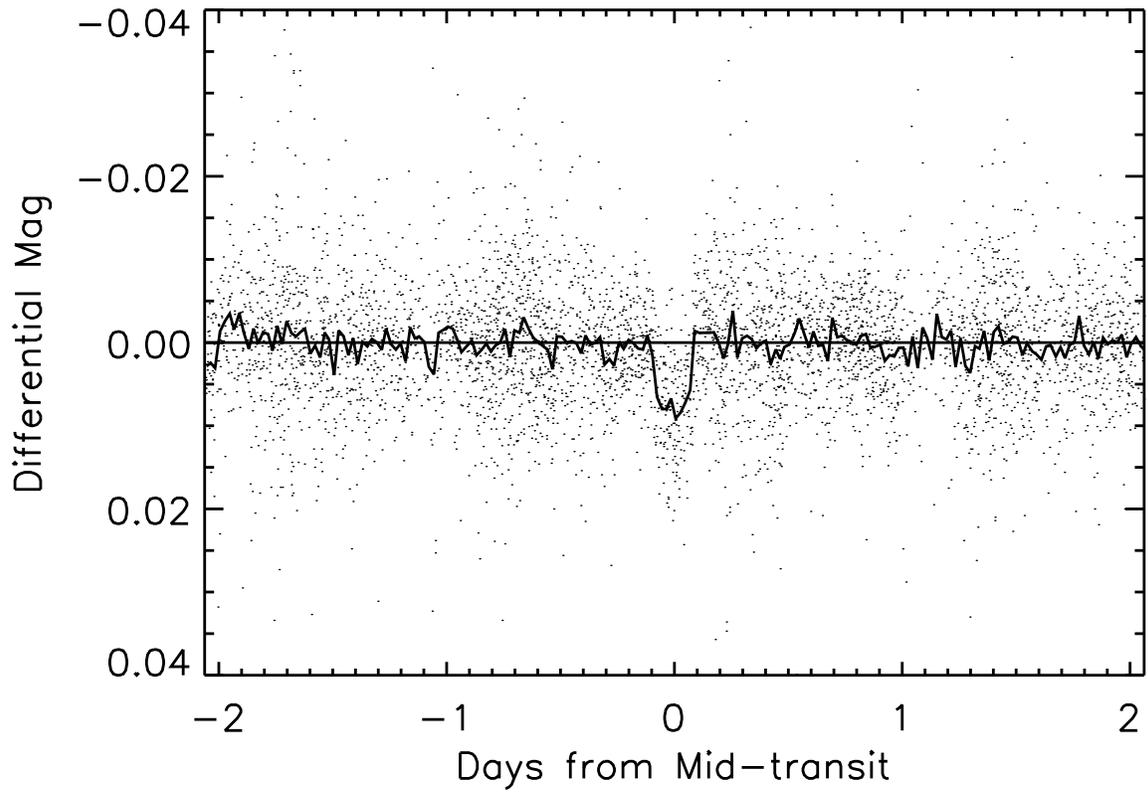}
\caption{
Photometry of \xon\ by the two XO cameras
over two seasons from Nov 2005 to Mar 2007 are shown wrapped and phased according
to the ephemeris of Equation \protect{\ref{eq:ephem}} and averaged
in 30-minute bins (line).
\label{fig:xolc}}
\end{figure}

In many ways \xon\ presented an excellent signature of a transiting
planet in the XO survey photometry.  Figure \ref{fig:finder} shows the
star is relatively isolated: our automated scripts had estimated that
86\% of the flux within the XO cameras' 75\arcsec-radius photometric
aperture was due to \xon.  The pixel centroid shifts of \xon\ are
consistent with the transit signal being intrinsic to \xon\ rather
than the result of a fainter eclipsing stellar binary within the
photometric aperture.  Making use of the TYCHO-2 and 2MASS catalogs,
the 19 mas yr$^{-1}$ proper motion combined with V$_{T}$=10.85 and
(V$_{T}$-J)=1.19 places \xon\ on the reduced proper motion diagram of
Gould \& Morgan (2003) within the $R_{\star}<1.25$ \Rsun\ cutoff,
although by the end of our analysis, we would derive a larger radius.
Transits were $\sim$1\% deep, flat-bottomed, and steep-edged, all
signs of a good candidate for a transiting planet (Figure \ref{fig:xolc};
Table \ref{table:photometry}).  Initially, the 4-hour
transit duration from the XO survey data was too long
given the initial G8V spectral type estimate for \xon\ from catalog broad band
photometry.  However, more accurate photometry and
spectroscopy yields an earlier, \sptype\ spectral type and
larger $R_{\star}$, which is consistent with the
4-hour transit duration.

The time line for XO observations covers $\sim$ 100 cycles of \xonb,
which provided an ephemeris sufficiently accurate for follow up photometric
observations to recover the transit.  The XO survey observed ingresses
of five separate transits and egresses of six transits, with only 
one transit observed in its entirety from ingress to egress.
With multiple transits
observed and with at least one pair separated by only one cycle, the
chance of misidentifying a longer-period orbit with a shorter-period
harmonic was small.

Extended Team\footnote{The XO Extended Team consists of volunteers who
provide photometric follow up of exoplanet candidates as described
by McCullough \& Burke (2007).}
time-series photometry in 2007 November confirmed
\xon\ is the variable star, verified and improved the transit's ephemeris
and shape, i.e. depth, duration, and abruptness of ingress and egress.
The Extended Team recorded time-series photometry of seven transits;
the data are in Table \ref{table:photometry}. 
A planetary transit has a nearly achromatic shape; many triple stars
do not.
The Extended Team observed
transits of \xonb\ once each in B, V, and I spectral bands and
multiple times in R band. The observed
shape of the transit is the same in each band, and
each observed depth equals the average depth (1.0\%) in R band, within
the typical uncertainty of 0.1\%.

All-sky photometric measurements were made on three dates (2008
January 10, 19 and 20) using a 0.35-meter telescope located in
Arizona in B, V, $\rm
R_c$, and $\rm I_c$ bands. Eight Landolt star fields and the \xon\
star field were observed, between airmasses of 1.18 to 1.3.
For the B and V bands a total of 54 Landolt stars were observed
multiple times per night; for the $\rm R_c$ and $\rm I_c$ bands 33
Landolt stars were observed multiple times.  The Landolt field
observations provide solutions to the color and airmass terms for each
night.  Applying the solution to the \xon\ field yields the standard
photometry as given in Table~\ref{table:allsky}.  The r.m.s. residuals of
the Landolt standard stars around the solution provide the
uncertainties in photometry.

\subsection{Scintillation-limited Photometry with a 1.8-m Telescope\label{sec:bestlc}}

With the Perkins 1.8-m telescope at Lowell Observatory, we obtained R band
time series photometry on 2008 January 20 (UT).  We
obtained five-second CCD exposures at 8-second cadence for six hours
of \xon\ and a single comparison star, for a total of 2448 exposures
(Figures \ref{fig:bestlc} and \ref{fig:inout}).
Incidentally, these observations with the PRISM instrument of
the \xonb\ transit followed those made by the same telescope and instrument earlier
the same night of a transit of XO-5b (Burke et al.\ 2008).  The PRISM sensor
is a cryogenically-cooled Fairchild CCD with 2048 pixels by 2048 pixels,
of which we used a 617 pixel by 144 pixel subarray, or 4\arcmin.01 by
0\arcmin.94 at 0\arcsec.39 per pixel, in order to reduce readout times
and thereby to increase the fraction of elapsed time collecting photons.
The system gain was 2.66 e$^-$ ADU$^{-1}$; the read noise was 7.8 e$^-$
r.m.s., and the detector has been demonstrated to be linear to within
$\pm$1\% for fewer than 1.0$\times 10^5$ electrons pixel$^{-1}$. The
shutter and filters are located near a re-imaged pupil, so any variations
from ideal performance will be identical across the focal plane and
removed by the differential photometry.  The telescope was intentionally
de-focused to produce annulus-shaped stellar images
in order to reduce the peak
irradiance on the CCD detector from the brightest star to
$\la 0.8 \times 10^{5}$ electrons pixel$^{-1}$ per 5-s exposure.  The de-focused
stellar images gradually increased in diameter from 3\arcsec\ to 6\arcsec\ 
as the telescope
tracked the star from airmass 1.1 to 2.1.  The telescope has an offset
guider which maintained the image position to $\sim 1$ pixel r.m.s. in
each axis, throughout the 6 hours of
observation, except for 10 minutes during which the images of the stars
drifted by 30 pixels (3 diameters of the de-focused stars) and re-stabilized
at a new position. 
The 10-minutes of drifting, from 0.063 to 0.070 days after mid-transit,
are indicated by a hash mark at bottom of Figure \ref{fig:bestlc}.

\begin{figure}
\plotone{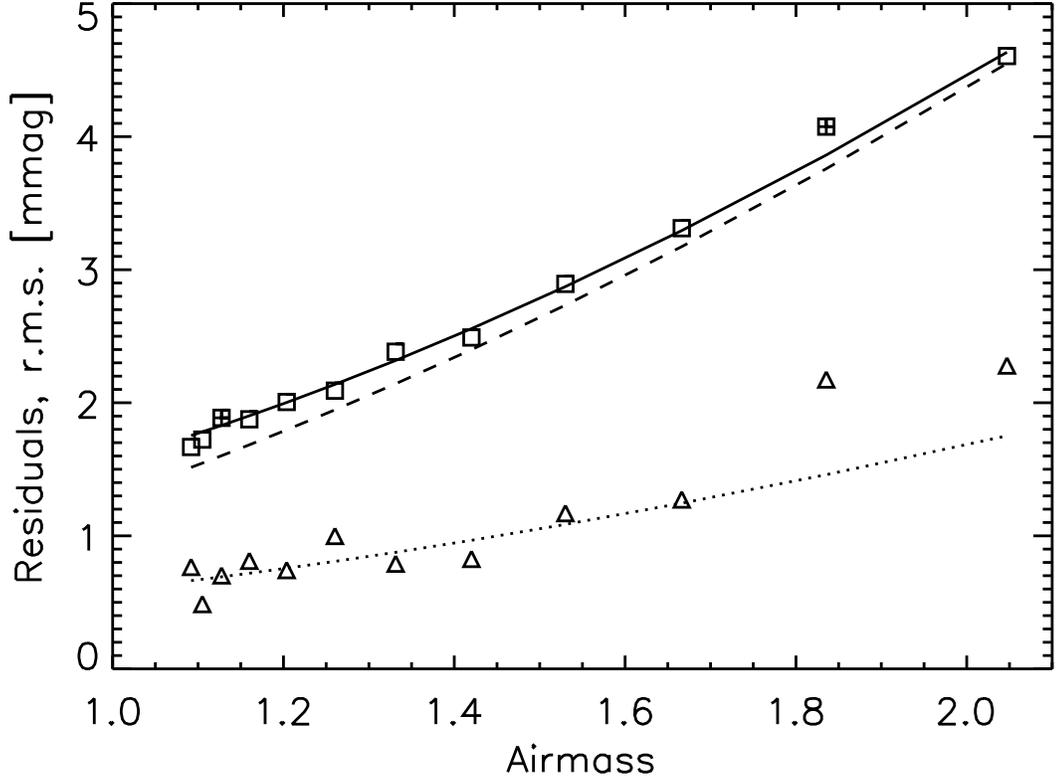}
\caption{Scintillation-limited photometry with the Perkins 1.8-m telescope in R band.  
The data (squares) are the standard deviations of $\sim 225$ samples within each of
twelve 30-minute intervals of differential photometry of the star \xon\ 
after a best-fit model to the transit of \xonb\ was subtracted. Data at the moments
of ingress and egress are indicated by $+$ symbols within the squares at airmasses
1.13 an 1.84, respectively.
The solid line is the quadrature sum of contributions from scintillation (dashed line)
and Poisson noise (0.9 mmag per 5-sec exposure).
The dashed line is proportional to the airmass raised to the power 1.75, as expected for
scintillation (see text).
The triangles are the same data averaged in 1-minute intervals, with 7 samples
per minute. The dotted line is $\sqrt{7}$ times less than the solid line.
\label{fig:scint}}
\end{figure}

A single comparison star, GSC 03793-01990 (labeled ``1" in
Figure \ref{fig:finder}),
was selected based on its proximity to \xon\ in location, brightness,
and color. With respect to \xon, the comparison star is 3\arcmin\ west,
0.21 mag brighter, and 0.04 mag redder in V-R$_C$ color.  The photometry
appears to be limited by scintillation (Figure \ref{fig:scint}).
In a 5-second exposure at low airmass, the camera collected 3.2 and
4.0 million photons from \xon\ and the comparison star, respectively,
implying a Poisson contribution to \xon's differential photometric
error of 0.9 mmag per exposure.  The noise increases with airmass,
indicating the Earth's atmosphere limits the photometric precision.
The increase in noise with airmass is predominately due to scintillation
($\propto$ airmass$^{1.75}$) and marginally due to fewer photons due
to the measured extinction of 0.12 mag
airmass$^{-1}$.  If we average every seven consecutive exposures into
samples with 1-min cadence, the r.m.s. of those samples is 0.6 mmag for
exposures near zenith, commensurate with the 1.8 mmag r.m.s. per exposure.
In Figure \ref{fig:scint} the observed r.m.s. noise is fit well by the
solid line, which corresponds to the quadrature sum of Poisson noise
from each star and uncorrelated scintillation of each star that is 25\%
larger than the nominal value given by Dravins et al.\ (1998, Eq. 10).
In the photometric noise model, Poisson and scintillation noise are both
$\sim\sqrt{2}$ larger than the noise contribution from a single source
since the differential light curve is calculated from the magnitude
difference between \xon\ and the comparison star.

\subsection{Spectroscopy\label{sec:spectroscopy}}

On 2007 November 20-21, we obtained cross-dispersed echelle spectra of
\xon\ with the 2dcoude spectrometer (Tull et al.\ 1995) on the 2.7-m Harlan
J.\ Smith (HJS) telescope at McDonald Observatory.  Thorium-argon lamp
spectra immediately preceding and following each R=60,000 stellar spectrum
yielded 0.1 \kms\ radial velocity precision. The \xon\ observations
were at similar zenith distance (34-35 degrees) both nights to improve
radial velocity precision.  These observations eliminated
a stellar or brown dwarf companion in an isolated binary
as the cause of the transits, leaving more complex systems (e.g.,
Mandushev et al.\ 2005) or a planetary companion as the only plausible
explanations. Section \ref{sec:sme} describes our analysis of the HJS spectra to
obtain stellar parameters.

Between 2007 December 21 and 2008 January 15 we obtained cross-dispersed
echelle spectra of \xon\ with the high-resolution spectrometer (Tull
1998) on the 11-m Hobby-Eberly Telescope (HET), also located at McDonald
Observatory. The R=63,000 spectra were obtained through an iodine gas
absorption cell to provide an accurate wavelength reference for each
observed spectrum. We obtained one spectrum per night for the first two
epochs to confirm the planetary signal, and then two spectra per night for
the remaining 7 epochs to improve radial velocity precision by $\sqrt 2$.
We processed each echelle spectrum from both telescopes
using the optimal extraction procedure
described in Hinkle et al.\ (2000). Section \ref{sec:rv} describes our analysis
of the HET spectra to measure radial velocity variations.

\section{Analysis\label{sec:analysis}}

\subsection{Ephemeris}

The heliocentric Julian date of minimum light (also mid-transit) is
\begin{equation}
t_{m.l.} = t_c + P \times E,
\label{eq:ephem}
\end{equation}
where E is an integer, $t_c = $\vjd$\pm$\ejd\ (HJD), and the orbital
period, $P =$\vperiod $\pm$\eperiod\ days.  We determine the epoch $t_c$
and its uncertainty using a Monte Carlo method (Section \ref{sec:lightcurve})
of modeling the
high-fidelity observations of the transit observed on Jan 20, 2008. We
determined the period from $t_c$ and the XO survey observations of
two ingresses and two egresses observed 2 years earlier.  Due to the
10-minute sampling and $\sim$1\% precision of the survey photometry, each
ingress or egress has a timing uncertainty of $\sim$10 minutes. With
four of them observed $\sim$180 cycles from $t_c$, the uncertainty
of the period is $\sim 10 {\rm min} / \sqrt{4} / 180$ = \eperiod\
days. 
Although the formal estimate of the period's uncertainty is far
smaller than required for any of the analysis reported here, future
planning of a time-critical observation, such as a sequence of spectra to
measure the Rossiter effect, one or more years from $t_c$ may benefit from
first verifying the ephemeris with additional precision photometry.

\subsection{Radial Velocities\label{sec:rv}}

We measured \xon\ radial velocities by modeling each HET spectrum
(obtained with an I$_2$ gas absorption cell in the light path) as the product
of a shifted stellar template spectrum multiplied by the known absorption
spectrum of the I$_2$ cell and then convolved with the line spread function
of the spectrograph.
To construct the stellar template, we scaled a high resolution
($\lambda/\Delta\lambda \approx 10^6$) solar spectrum (Wallace,
Hinkle, \& Livingston 1998) to match the observed line depths in
each wavelength interval and then convolved with a rotational
broadening kernel that includes the effects of stellar limb-darkening.
We obtained an FTS spectrum of the HET I$_2$ cell (Cochran 2000)
from the National Solar Observatory online archive.

\begin{figure}
\plotone{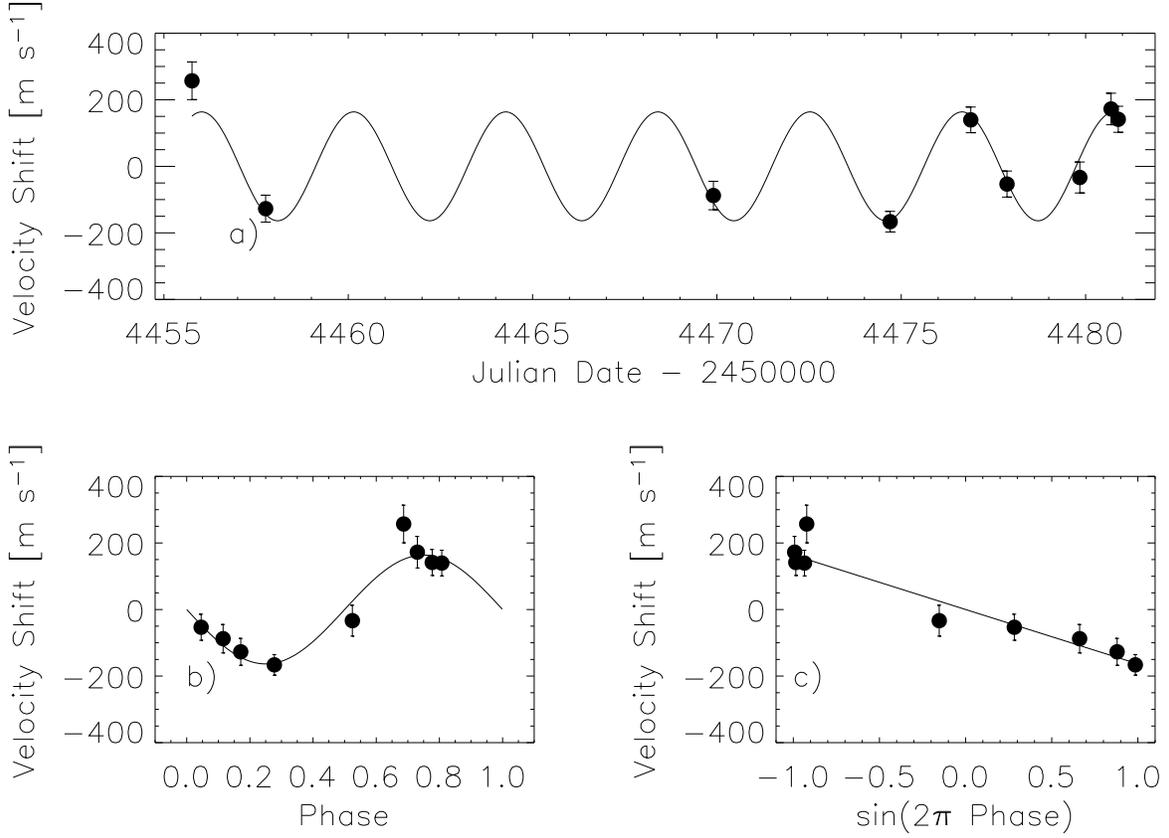}
\caption{a) The radial velocity of \xon\ oscillates sinusoidally with a
semi-amplitude K = \vrvK$\pm$\ervK\ \mps.
b) The period and phase of the radial velocities were fixed at values determined
by the transits. The mean stellar radial velocity with respect to the
solar system's barycenter has been subtracted. 
c) In this representation of
the data, a circular orbit yields a straight line of slope $-$K.
\label{fig:rviodine}}
\end{figure}

For each wavelength interval, the model parameters are a continuum
scale factor, an exponent that scales line depth, the iodine wavelength
shift, the stellar radial velocity, and slight deviations from a nominal
Gaussian line spread function.
Using downhill-simplex $\chi^2$ minimization, we adjusted these free
parameters to fit $\sim$32 separate 1.5 nm wavelength intervals with
significant I$_2$ absorption (521--570 nm). 
For each observed spectrum, we compute the mean radial velocity and
adopt the standard deviation divided by $\sqrt{32}$ as the uncertainty
in the mean. 
The radial velocities, transformed to the barycentric frame of the solar
system, are in Table \ref{table:rv} and Figure \ref{fig:rviodine}.
We phased the radial velocities to the ephemeris of the transits,
assumed a circular orbit, and determined the maximum likelihood radial
velocity semi-amplitude K = \vrvK$\pm$\ervK\ \mps.

Precise line bisector measurements (e.g., Torres et al.\ 2005; Johns--Krull
et al.\ 2008) can be used to detect triple star systems that produce shallow
transits that may be misinterpreted as planetary transits.
For each of our HET spectra, we measured the mean bisector span of
stellar absorption lines free of I$_2$ and telluric absorption.
We find no significant correlation with measured radial velocities, but the
significance of this result is limited by the relatively low signal noise ratio
of our spectra.

\subsection{Stellar Properties from Spectroscopy and Isochrone Analysis\label{sec:sme}}

We used the SME package (Valenti \& Piskunov 1996) to fit each observed
HJS spectrum (Section \ref{sec:spectroscopy}) with a synthetic spectrum,
adopting the same wavelength intervals (5150-5200 and 6000-6200 \AA),
line data, atmospheres, and post facto parameter adjustments as Valenti
\& Fischer (2005).
Table \ref{table:spec} lists the resulting stellar parameters for \xon:
  effective temperature (\Teff),
  logarithm of the gravity (\Logg),
  metallicity (\MonH),
  projected rotational velocity (\Vsini),
  and
  logarithm of the abundances of Na, Si, Ti, Fe, and Ni relative to solar
    (\NaH, \SiH, \TiH, \FeH, and \NiH).
Our \MonH\ parameter is an abundance scale factor for elements other than
Na, Si, Ti, Fe, and Ni, so it is not equivalent to standard metallicity.
The 1$\sigma$ uncertainty for each parameter is listed in the ``Uncer''
column.

We ran the spectroscopic analysis four times, initially allowing \Logg\
to be a free parameter (``Run 1'') and then fixing it at three specific
values.
Valenti \& Fischer (2005) gravities for stars cooler than \xon\ are
constrained almost entirely by the collisional damping wings of the \Mgb\
triplet lines.
For stars as warm as \xon, these damping wings become relatively weak,
providing less of a constraint on stellar gravity.
Fortunately, the shape of a precise transit light curve provides a strong
independent constraint on stellar gravity (e.g., Winn et al.\ 2008).
In Table \ref{table:spec}, ``Run 3'' (in bold) is our preferred solution,
as it assumes the value of \Logg\ favored by the transit light curve
analysis (Section \ref{sec:lightcurve}).
Tabulated results for the other three runs illustrate in a useful format
the covariance of stellar parameters, subject to our spectroscopic constraint.
Reduced $\chi^2$ for each spectroscopic fit are listed in the \rchisq\ row. 
SME users should note that Run 1 did not converge to the absolute minimum
value \rchisq\ because the Levenberg-Marquardt algorithm used in SME can
have some difficulty following shallow valleys in the $\chi^2$ surface
that are not aligned with parameter axes.
 
\begin{figure}
\plotone{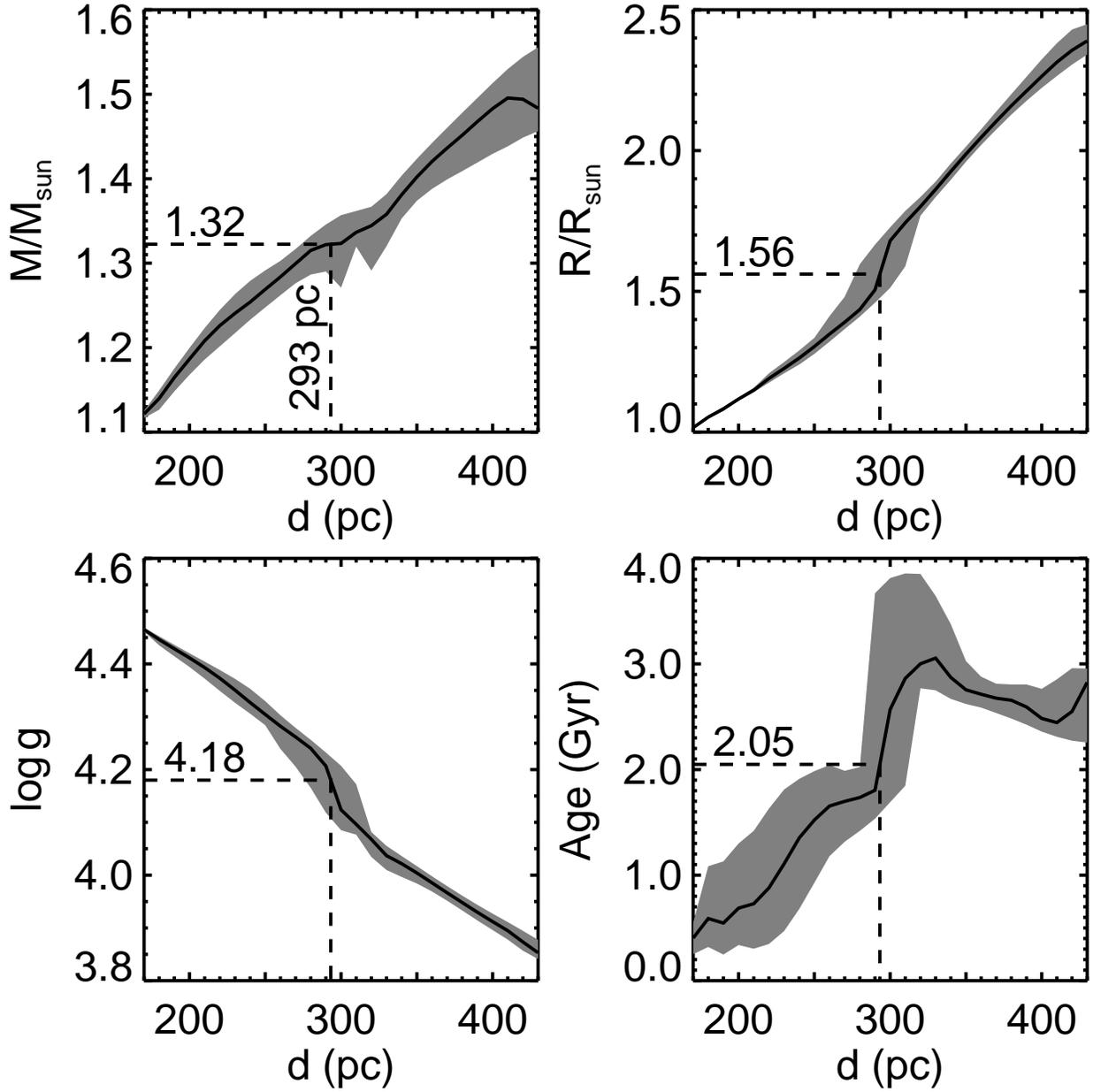}
\caption{
Credible intervals for \xon\ stellar properties as a function of
assumed distance, based on the ``Run 3'' isochrone analysis with
$\Logg=\vlogg$ from the light curve analysis.
\label{fig:iso}}
\end{figure}

To obtain stellar mass (\Mstar), radius (\Rstar), and age, we interpreted our
spectroscopic and photometric results using Yonsei--Yale (\Ytwo) isochrones
(Demarque et al.\ 2004).
We calculated a bolometric correction and interpolated the isochrones
using the procedure described in Valenti \& Fischer (2005), except that
for \xon\ we were forced to assume a sequence of possible distances
(180 to 430 pc in steps of 10 pc).
Setting the isochrone gravity equal to the gravity used in the SME analysis
selects a preferred distance and hence preferred values for stellar mass,
radius and age. 
Figure \ref{fig:iso} shows credible parameter intervals for Run 3, which
yielded a preferred radius ($1.56\Rsun$) nearly identical to the radius
favored by the light curve analysis ($\vRs\Rsun$).
Using the parameter relationships in Table \ref{table:spec}, we can
translate the $\eRs\Rsun$ uncertainty in \Rstar\ (from the light curve
analysis) into corresponding uncertainties in distance, stellar gravity,
stellar mass, and age. In Table \ref{table:spec} we adopt uncertainties
three times these nominal values to account crudely for possible systematic
errors in our analysis.

\subsection{Light Curve Modeling\label{sec:lightcurve}}

\begin{figure}
\plotone{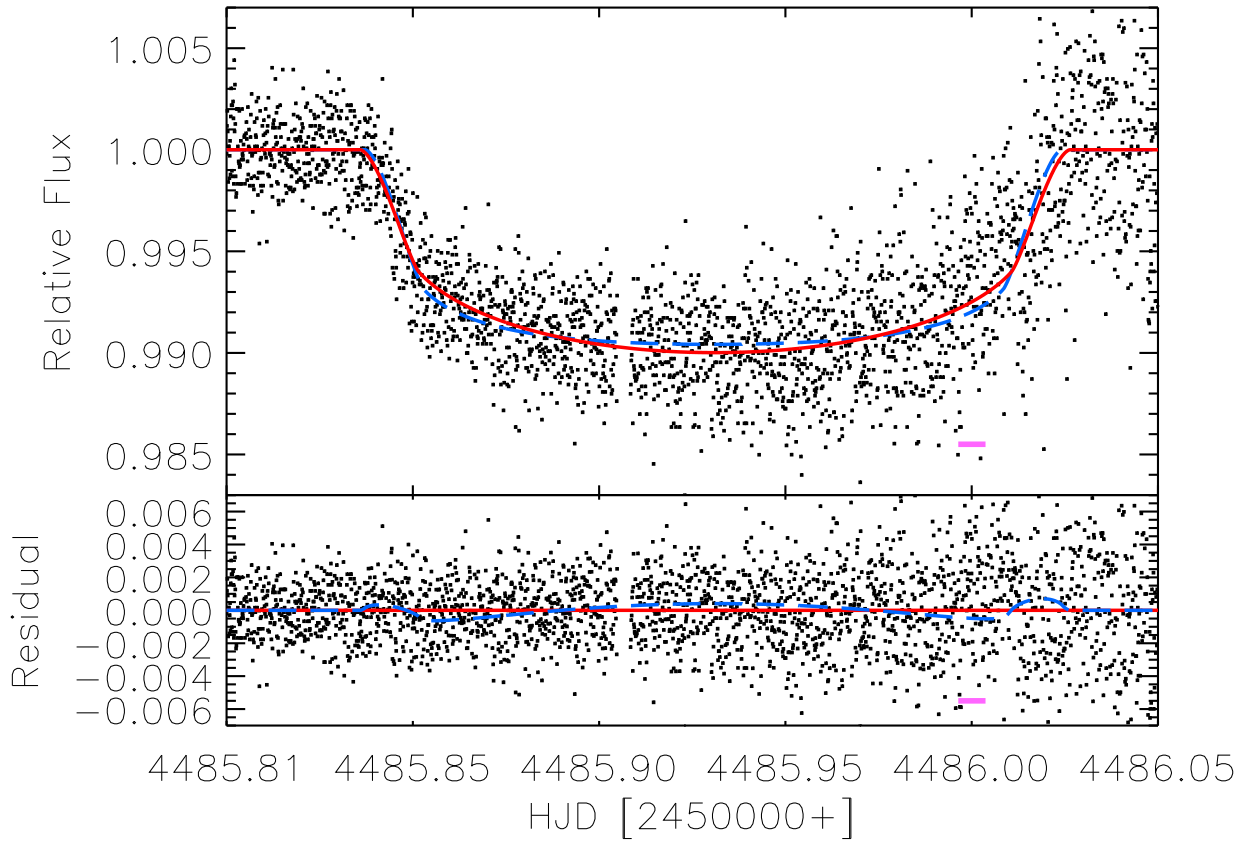}
\caption{{\it Top}: Time-series R-band photometry of \xon\ as observed with the Lowell 1.8-m
Perkins telescope and PRISM camera during the Jan 20, 2008 transit of \xonb.
The data (2448 points), the best-fitting model in a $\chi^{2}$ sense from the MCMC analysis (solid line), and the best-fitting model in a $\chi^{2}$ sense with limb darkening coefficients fixed at the R-band theoretical values of Claret (2000) (dashed line). {\it Bottom}: Residual from the best-fit model from the MCMC analysis (points) and difference between models with variable and fixed to the theoretical limb darkening coefficients (dashed line). 
\label{fig:bestlc}}
\end{figure}

In order to determine the physical parameters of the star and the
transiting planet (Table \ref{table:planet}),
we modeled the high-fidelity transit light curve (Figure \ref{fig:bestlc})
using the transit model of Mandel \& Agol (2002) and the Markov Chain Monte Carlo
(MCMC) methodology.  Because we used
the same MCMC procedure as Burke et al.\ (2007; 2008), in this paper 
we describe the assumptions and results but do not describe the MCMC procedure itself.
For simplicity, we assume a circular orbit;
Section \ref{sec:discussion} discusses implications
of the circular-orbit assumption being invalid.

To fully determine the system parameters from the transit light curve,
we adopt an informative prior for $M_{\star}=1.34\pm 0.08$ \Msun.
This initial estimate for $M_{\star}$ comes from the SME isochrone analysis
(\S~\ref{sec:sme}).  The uncertainty in the prior for $M_{\star}$
conservatively agrees with the typical uncertainties in $M_{\star}$
for the homogeneous analysis of other known transiting planets of
Torres et al.\ (2008).  Analysis of the transit light curve provides a
more precise estimate of log$g$ than the spectroscopic determination.
One iteration of the SME analysis with log$g$ fixed to this more
precise estimate, the resulting variation to $M_{\star}$ was not
significant for our given prior on $M_{\star}$. The prior on each of the
other parameters, in particular $R_{\star}$ and $R_{\rm p}$, is uniform.
The noise in the light curve averages down as expected on one minute
times scales thus we assume Gaussian indendent noise for the
Likelihood function.  For the uncertainties associated with each photometric measure of the light curve,
we use the analytic model of Poisson noise and scintillation noise (Section
\ref{sec:bestlc}; Figure \ref{fig:scint});
we note that $\sigma_i = 1.9$ mmag at ingress and $\sigma_e = 4.1$ mmag at egress.
The free parameters in the MCMC fit are $M_{\star}$, $R_{\star}$,
$\rho=R_{\rm p}/R_{\star}$, $\tau$, $t_{o}$, $u_{1}$, $u_{2}$, and
$zpt$, where $\tau$ is the total transit duration from 1$^{\rm st}$ to
4$^{\rm th}$ contact, $t_{o}$ is the transit timing midpoint offset
from an initial ephemeris, $u_{1}$ and $u_{2}$ are the limb darkening
coefficients for a quadratic law, and $zpt$ is the flux ratio zero
point for the differential light curve.

Table~\ref{table:spec} and Table~\ref{table:planet} show the resulting
parameters for \xon\ and \xonb, respectively.  The median of the MCMC
samples provides a robust (hereafter ``best'') estimate of each parameter
and the uncertainties are the
symmetrical confidence interval containing 68\% of the samples.  The
maximum-likelihood model, i.e. in a $\chi^{2}$ sense, is
the solid line in Figure~\ref{fig:bestlc}, which also
shows the data residuals around the maximum-likelihood model.  

The best estimate for the limb darkening coefficients ($u_{1}=0.61\pm
0.1$ and $u_{2}=0.10\pm 0.23$ are significantly different than the
theoretical R-band limb darkening coefficients from Claret et
al. (2000) for a star with the properties of \xon\ ($u_{1}=0.24$ and
$u_{2}=0.38$).  The theoretical B-band limb darkening coefficients
($u_{1}=0.49$ and $u_{2}=0.29$) are closer in agreement to what is
measured.  To investigate this difference we perform a $\chi^{2}$ fit
to the Perkins transit light curve fixing the limb darkening
parameters at their theoretically expected value in the R-band
($M_{\star}=$\vMs\ also held fixed).  The dashed line in
Figure~\ref{fig:bestlc} shows the resulting transit model.  The
theoretical limb darkening coefficients result in steeper ingress and
egress with a slightly shallower mid transit flux level.  
In Figure~\ref{fig:bestlc}, the plot
of residuals is with respect to the model with the 
limb darkening coefficients treated as free parameters
in the MCMC analysis, thus the solid line is zero by definition and
the dashed line equals the difference due to fixing the limb darkening
coefficients at their theoretical values.

\begin{figure}
\epsscale{0.9}
\plotone{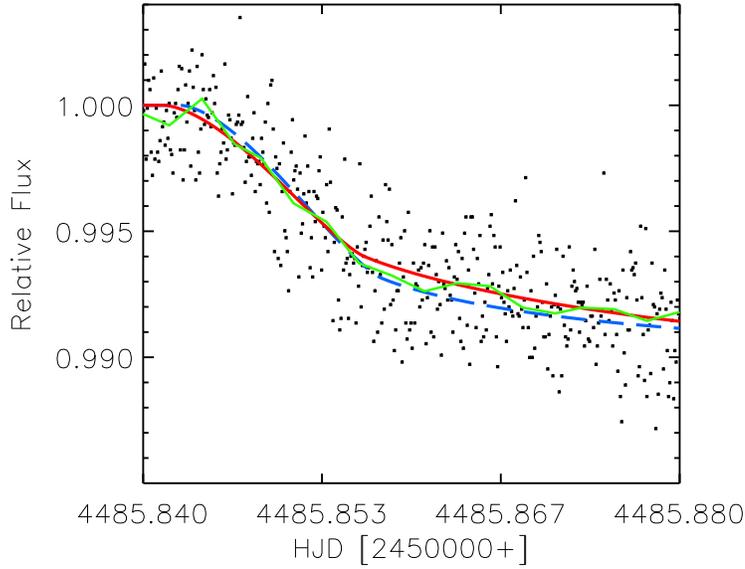}
\plotone{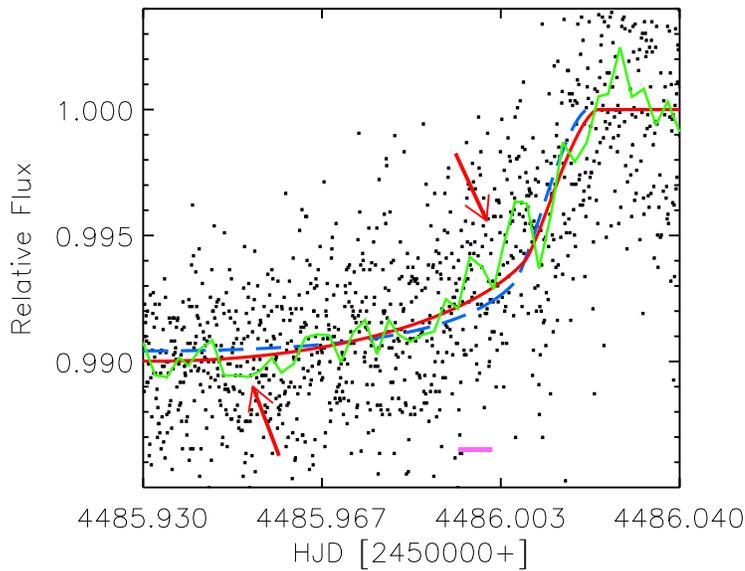}
\caption{Detail of ingress (upper) and egress (lower) of the Lowell
1.8-m Perkins light curve (points) along with data in bins (jagged
solid line).  Best-fitting transit model in a $\chi^{2}$ sense from the
MCMC analysis with limb darkening coefficients as free parameters (smooth
solid line). Best-fitting transit model in a $\chi^{2}$ sense with limb
darkening coefficients fixed at the theoretical values from Claret (2000)
(smooth dashed line).  Two epochs show the largest amplitude difference
between data and transit models (arrows).
\label{fig:inout}}
\epsscale{1.0}
\end{figure}

Formally, under the assumption of Gaussian independent noise, the
theoretical limb darkening coefficients result in $\Delta\chi^{2}=48$
worse fit at $>$ 6-$\sigma$ for 2 degrees of freedom.  To investigate
the reliability of these models, we zoom in on the egress portion of
the light curve in Figure~\ref{fig:inout}, where the largest amplitude
difference between the models occurs and also the largest photometric
noise.  The smooth solid and dashed lines show the best-fit model with
varying and fixed limb darkening coefficients, respectively.  The
jagged solid line shows binned data to improve the
visibility of the light curve.  In our data the two most significant
excursions from either model are labeled with
arrows in Figure~\ref{fig:inout}.  We consider two possibilities: either the
noise model is correct and we are measuring limb darkening
coefficients significantly in conflict with the theoretical ones, or
the noise model overestimates the significance of these two excursions
of the light curve, causing the limb darkening
coefficients to compensate for systematic errors in the light curve.  We
consider the latter is the case for the excursion near
third contact, because during that time, the stars' positions were
shifting on the detector (see \S~\ref{sec:bestlc}). On the other hand, other
systematic deviations do not correlate with variations in external parameters;
an example prior to third contact is indicated also in Figure \ref{fig:bestlc}.
From analysis of ground-based and space-based light curves, Southworth (2008)
concludes that if limb-darkening coefficients are not included as fitted
parameters, uncertainties in other parameters may be underestimated, because the
highest-quality light curves (e.g. spectrophotometric, space-based
observations of HD 209458b)
show significant differences between theoretically-predicted and
observationally-derived limb-darkening coefficients.
We concur with Southworth (2008) that multiple observations may
be useful to discern limitations of a single light curve.
The choice of limb darkening profile
does not significantly affect $R_{\star}$ or the planetary
properties in the case of \xonb.  The difference in parameters,
$\Delta R_{\star}=0.07\pm 0.05$ \Rsun,
$\Delta \rho=-0.0017\pm 0.0013$,
$\Delta R_{\rm p}=0.03\pm 0.05$ \Rjup,
$\Delta t_{o}=0.4\pm 0.6$ mn, are in the sense of MCMC analysis minus fixed limb darkening coefficients and the uncertainty is the uncertainty in the parameter from the MCMC analysis only.

\section{Discussion and Conclusion\label{sec:discussion}}

Table \ref{table:compare} compares physical characteristics of
\xon\ and two similar transiting systems,
HAT--P--4 (Kovacs et al. 2007) and HAT--P--6 (Noyes et al. 2008).
The HAT systems were selected for comparison because their
stars are similar to \xon.
The relatively large duty cycles of the transits of all three
planets imply low mean densities for
the host stars, a fact confirmed by the SME analysis of their spectra.
The three planets radii are similar but their masses differ by
as much as a factor of 2.5.

The mass of \xonb\ (\vMp\ \Mjup) places it at the 
the margin of the bulk of the distribution of planetary masses
for planets with orbital periods between 3 and 5 days.
For such planets, the distribution declines rapidly with mass for $M_p \ga $1 \Mjup,
and the upper $\sim$10\% of the distribution of gas giants
(4 of $\sim$40 reported) is spread broadly over the range from 1 to 12 \Mjup.
The latter approximate description is true for both the set of planets
that transit, for which the mass is known, and the set of
planets that do not, for which a minimum mass is known.
The two sets are similar in number, $\sim 20$ planets each. 
That the distribution has a sharp decline at $\sim$1 \Mjup\ (or between 1 and 2 \Mjup)
is convincing, but the shape of the tail of the distribution for $M_p \ga $1 \Mjup\ 
is poorly determined for this period range, due to the small number of planets.
\xonb\ increases that small number by one.

Analysis of transiting systems yields orbital inclinations and stellar radii.
These quantities may be combined with \Vsini\ to estimate stellar rotation
periods,
\begin{equation}
\Pstar={2\pi\phi\Rstar\over\Vsini},
\end{equation}
where $\phi$ is a factor of order unity that accounts for differential
rotation and any systematic errors in \Vsini.
SME yielded $\Vsini=1.7$ \kps\ for the Sun (Valenti \& Piskunov 1995),
which rotates every 24 days at the equator and every 30 days at the pole.
Adopting 27 days as the characteristic rotation period of the Sun implies
$\phi=0.91$.
With this value of $\phi$, the rotation periods of the three warmest stars
known to host transiting planets are 8.1, 3.4, 7.7 days for XO--4, XO--3,
and HAT--P--6, respectively.
Adopting $\Rstar=1.38\Rsun$ from Winn et al.\ (2008) for XO--3, these stellar
rotation periods are 2.0, 1.1, and 2.0 times the planetary orbital periods.
XO--3b is massive enough that the star (or at least the convective envelope) 
may have been forced into synchronous rotation.
\xonb\ and HAT--P--6b have orbital periods that are twice the stellar rotation
period, suggesting that their orbits may be affected by resonant interactions
with their rotating host stars.
On the other hand, planets (e.g., HAT--P--2b and HAT--P--7b) orbiting
slightly cooler stars do not have orbital periods that are small multiples
of the stellar rotation period.
Perhaps Jupiter mass planets are only able to interact effectively with
stars that have very shallow surface convection zones.
The discovery of additional planets that transit warm F dwarfs will test whether
resonant interactions can affect planetary orbital periods or stellar
rotation periods.
One issue to keep in mind, however, is a possible selection effect against
stars with large \Vsini\ because broad lines make
difficult measuring radial velocity variations and the process
of discriminating planets from triple star systems.

We emphasize that this analysis assumed the orbit of \xonb\ is circular,
i.e. $e=0$.
If that assumption is false, i.e.
$e > 0$, logically the conclusions are invalid and all the
derived physical parameters will change accordingly.
The planetary mass is proportional to $K \sqrt{(1-e^2)}$
(Hilditch 2001; Equation 2.53), so only an eccentricity 
$\ga 0.5$ would change the planetary mass estimate significantly
compared to its fractional uncertainty ($\sim$10\%).
From the measured depth of the transit, the planetary radius is proportional
to the stellar radius, and the latter is proportional to the transverse
velocity of the planet at transit. The latter velocity depends on the argument
of periastron but is bound by its values at periastron and apasteron, i.e.
$1\pm e$ times its value for a circular orbit of the same period.
Hence an eccentricity
$e \ga 0.035$ could change the planetary radius estimate by $\ga 1$-$\sigma$.
Interestingly, the {\it amplitudes} of the radial velocity
curve and the light curve readily reveal the {\it ratios} of
planetary-to-stellar masses and radii, but the detailed {\it shapes}
of \underline{both} curves are required to measure precisely the radii
distinctly.

In addition to assuming $e = 0$, one might also assume that the
impact parameter $b = 0$, i.e. that the planet's path crosses
the center of the star. Although there is no physical
justification for such an assumption, it can be helpful in bounding the limits
of the derived physical parameters.
If both $e = 0$ and $b = 0$, the density of the star $\rho_*(e=0;b=0)$ is
determined by the orbital period and the
duration of the transit. In that case, the derived radius of the star
$R_*(e=0;b=0)$ is simply proportional to $M_*^{-1/3}$, where the mass
of the star, $M_*$ is estimated from the spectra and the isochrone analysis. 
The density $\rho_*(e=0;b=0)$ is an upper limit,
i.e. $\rho_*(e=0;b=0) \ge \rho_*(e=0;b\ge 0)$. 

The scintillation-limited, high-cadence photometry of \xon\ presented here
has a small r.m.s. per unit time, 0.6 mmag min$^{-1}$, at small airmass and
during planetary ingress, but which increases with airmass to
$\ga 2$ mmag min$^{-1}$ at airmass $\ga 2$, during planetary egress. 
To improve confidence in the parameters derived for the \xon\ system, 
additional high-fidelity time-series photometry would be beneficial.
Observations above the Earth's atmosphere are not limited by scintillation
and can more nearly achieve Poisson-limited results, which for \xon\ 
with a comparable spectral band and optical throughput,
would be $\sim 0.2$ mmag min$^{-1}$ if operational overheads can be made negligible.
The declination of \xon\ places it within the continuous viewing zone of
the Hubble Space Telescope (\hst), which permits observation without
interruption caused by occultation by the Earth. The latter circumstance
may enhance \xon's potential for precision spectrophotometry, because
those gaps and issues associated with them potentially may
cause systematic errors in precision time series obtained with \hst.

\acknowledgments

We thank the staff of the University of HawaiÔi, Institute for Astronomy, Haleakala
Observatories. We especially thank J. Heasley, M. Maberry, and R. Ratkowski
for assistance in operations on Maui. We thank the staffs of McDonald Observatory
and Lowell Observatory.

R. Bissinger, P. Howell, F. Mallia, G. Masi, K. Richardson, J. G., C. N. F., B. L. G.,
M. F., E. G.-M., and T. M. observe for the XO Extended Team.
Radek Poleski assisted with extrasolar planet candidate identification.
Lisa Prato and
Naved Mahmud assisted with observing at McDonald Observatory.
D. Bell transmitted HET data to us.

This research made use of the Royal Beowulf cluster at STScI;
the SIMBAD database, operated at CDS, Strasbourg, France;
data products from the Two Micron All Sky Survey (2MASS),
the Digitized Sky Survey (DSS),
and The Amateur Sky Survey (TASS);
source code for transit light-curves (Mandel \& Agol 2002);
and community access to the HET.
XO is funded primarily by NASA Origins of Solar Systems grant NNG06GG92G,
with additional funding from the STScI Director's Discretionary Fund.

\clearpage
\begin{deluxetable}{cccccc}
\tabletypesize{\small}
\tablewidth{0pt}
\tablecaption{{\rm Photometry}}
\startdata
\hline
\hline
Star${\rm ^a}$	&B 	&V	&${\rm R_C}$	&${\rm I_C}$ &V-${\rm R_C}$\\
\hline
\xon\ 	&11.240	&10.674 &10.324	&10.057 & 0.350 \\
   1 	&11.217 &10.503 &10.109 & 9.758 & 0.394 \\
   2 	&11.853 &10.329 & 9.482 & 8.748 & 0.847 \\
   3    &12.030 &11.613 &11.358 &11.084 & 0.255 \\
   4    &13.185 &12.526 &12.184 &11.798 & 0.342 \\
   5    &13.018 &12.314 &11.938 &11.560 & 0.376 \\
   6    &12.555 &11.475 &10.904 &10.390 & 0.571 \\
   7    &13.828 &13.121 &12.663 &12.297 & 0.458 \\
   8    &14.358 &13.787 &13.408 &13.073 & 0.378 \\
   9    &14.833 &13.930 &13.362 &12.834 & 0.568 \\
\enddata
\tablenotetext{a}{
Stars are identified in Figure \ref{fig:finder}. 
The 1-$\sigma$ uncertainties are 0.029, 0.019, 0.008, and 0.023 mag
for B, V, ${\rm R_C}$, and ${\rm I_C}$ respectively.
The 2MASS magnitudes for \xon\ are 9.667, 9.476, 9.406
for J, H, and K$_s$ respectively (Skrutskie et al.\ 2006).}
\label{table:allsky}
\end{deluxetable}

\begin{deluxetable}{cccccc}
\tabletypesize{\small}
\tablewidth{0pt}
\tablecaption{{\rm Time-Series Photometry$^a$}}
\startdata
\hline
\hline
HJD	& Brightness & Uncertainty	&Filter$^b$	& N$^c$ & Observer$^d$ \\
   	& [mag]      & (1-$\sigma$) [mag] &     &       &          \\
\hline
2453691.10864 & 0.0005 &0.0096 &   W  &1 & XO \\
2453691.10889 &-0.0052 &0.0088 &   W  &1 & XO \\
2453691.11645 & 0.0087 &0.0095 &   W  &1 & XO \\
2453691.11597 &-0.0043 &0.0088 &   W  &1 & XO \\
2453691.12256 &-0.0032 &0.0094 &   W  &1 & XO \\
\enddata
\\
\tablenotetext{a}{The entire table is in the electronic edition.
The printed edition contains only a sample to establish the format.}
\tablenotetext{b}{Standard filters, except W = wide, 400-700 nm.}
\tablenotetext{c}{Average of N measurements.}
\tablenotetext{d}{Observer initials, except XO is the XO cameras.}
\label{table:photometry}
\end{deluxetable}

\begin{deluxetable}{ccc}
\tabletypesize{\small}
\tablewidth{0pt}
\tablecaption{{\rm Radial Velocity Shifts}}
\startdata
\hline
\hline
Julian Date & Radial Velocity &  Uncertainty 	\\
            &  Shift [\mps] &  (1 $\sigma$) [\mps]  \\
\hline
2454455.7622  &    257   &   57 \\
2454457.7597  &   -127   &   40 \\
2454469.9042  &    -88   &   43 \\
2454474.6989  &   -166   &   31 \\
2454476.8855  &    140   &   39 \\
2454477.8677  &    -53   &   39 \\
2454479.8429  &    -34   &   46 \\
2454480.6910  &    172   &   47 \\
2454480.8856  &    141   &   39 \\
\enddata
\label{table:rv}
\end{deluxetable}

\begin{deluxetable}{lccccc}
\tablecaption{Stellar Properties of \xon\label{table:spec}}
\tablewidth{0 pt}
\tablehead{
  \colhead{Parameter}      &
  \colhead{Run 1}          &
  \colhead{Run 2}          &
  \colhead{\textbf{Run 3}} &
  \colhead{Run 4}          &
  \colhead{Uncer}
  }
\startdata
\multicolumn{6}{l}{\textit{Spectroscopic Analysis:}}\\
\Teff\ (K)     & 6249  &  6349 & \textbf{ 6397} &  6491 & 70      \\
\Logg\ (cgs)   & 3.98  &  4.09\tablenotemark{a}
                               & \textbf{ 4.18\tablenotemark{a}}
                                                &  4.30\tablenotemark{a}
                                                        & \elogg\tablenotemark{b}  \\
\MonH          & -0.05 & -0.02 & \textbf{-0.02} &  0.03 & 0.05    \\
\Vsini\ (\kms) &  9.0  &  8.7  & \textbf{ 8.8 } &  8.5  & 0.5     \\
\NaH           & -0.12 & -0.05 & \textbf{-0.02} &  0.07 & 0.2     \\
\SiH           & -0.02 &  0.01 & \textbf{ 0.03} &  0.04 & 0.02    \\
\TiH           & -0.01 &  0.00 & \textbf{-0.02} &  0.02 & 0.07    \\
\FeH           & -0.09 & -0.08 & \textbf{-0.04} & -0.03 & 0.03    \\
\NiH           & -0.21 & -0.04 & \textbf{-0.02} &  0.05 & 0.05    \\
\rchisq        &  1.82 &  1.77 & \textbf{ 1.77} &  1.81 & \nodata \\[8pt]
\multicolumn{6}{l}{\textit{Isochrone Analysis:}}\\
$d$ (pc)       &  343  &  312  & \textbf{ 293 } &  257  & \eDs\tablenotemark{b}  \\
$\Mstar/\Msun$ &  1.34 &  1.33 & \textbf{ 1.32} &  1.29 & \eMs\tablenotemark{b}  \\
$\Rstar/\Rsun$ &  1.96 &  1.76 & \textbf{ 1.56} &  1.33 & 0.05\tablenotemark{c} \\
Age (Gyr)      &  3.5  &  2.9  & \textbf{ 2.1 } &  1.3  & \eAge\tablenotemark{b} \\
\enddata
\tablenotetext{a}{Value of \Logg\ was fixed during the SME analysis.}
\tablenotetext{b}{Three times the uncertainty obtained by propagating $0.05\Rsun$
  uncertainty in \Rstar.}
\tablenotetext{c}{Uncertainty from light curve analysis.}
\end{deluxetable}

\begin{deluxetable}{lcl}
\tabletypesize{\small}
\tablecaption{{\rm The Planet \xonb}}
\tablewidth{0pt}
\tablehead{
  \colhead{Parameter}               &
  \colhead{Value$^a$}               &
  \colhead{Notes}               
}
\startdata
$P $\dotfill 			& \vperiod$\pm$\eperiod\ d 		& Period \\
$t_c$\dotfill			& \vjd$\pm$\ejd\ HJD 			& Transit midpoint\\
$e$(assumed)\dotfill 				& 0					& Eccentricity \\
$K $\dotfill 				& \vrvK$\pm\ervK$\ \mps 		& \\
$M_p $\dotfill 			& \vMp$\pm$\eMp\ \Mjup		 	& Mass\\
$R_p $\dotfill 			& \vRp$\pm$\eRp\ \Rjup			& Radius\\
$a $\dotfill 		& \vap$\pm$\eap\ A.U. 			& Semi-major axis\\
$a/R_* $\dotfill 		& \vaoRs$\pm$\eaoRs\ 			& \\
$R_p/R_* $\dotfill 		& \vRpoRs$\pm$\eRpoRs\  			& \\
$i $\dotfill 		& \vincl$\pm$\eincl\ deg		& Inclination\\
$b $\dotfill 				& \vbp$\pm$\ebp\ 	& Impact parameter\\
\enddata
\tablenotetext{a}{\Rjup\ = 71492 km; \Mjup\ = 1.8988e27 kg}
\label{table:planet}
\end{deluxetable}

\begin{deluxetable}{lcccc}
\tablecaption{Comparison of Three Transiting Systems\tablenotemark{a}\label{table:compare}}
\tablewidth{0 pt}
\tablehead{
  \colhead{Parameter}      &
  \colhead{HAT--P--4}          &
  \colhead{HAT--P--6}          &
  \colhead{\xon}          &
  \colhead{Uncer\tablenotemark{b}}
  }
\startdata
\multicolumn{5}{l}{\textit{Stars:}}\\
\Teff\ (K)     & 5860  & 6570  &  6397 &  70      \\
\Logg\ (cgs)   & 4.14  & 4.22  &  4.18 &  \elogg  \\
\Vsini\ (\kms) & 5.5   & 8.7   &  8.8  &  0.5     \\
\FeH           & +0.24 & -0.13 & -0.04 &  0.03    \\
Distance, (pc)       & 310   & 260   &  293  &  \eDs \\
Mass, (\Msun) & 1.26  & 1.29  &  1.32 &  \eMs \\
Radius, (\Rsun) & 1.59  & 1.46  &  1.56 &  0.05 \\
Age (Gyr)      & 4.2   & 2.3   &  2.1  &  \eAge \\[8pt]
\multicolumn{5}{l}{\textit{Planets:}}\\
Period (d)     & 3.06\tablenotemark{c}  & 3.85\tablenotemark{c}  &  4.13\tablenotemark{c} & n/a\tablenotemark{c} \\
Mass (\Mjup) & 0.68  & 1.06\tablenotemark{c} &  \vMp &  \eMp  \\
Radius (\Rjup) & 1.27  & 1.33  &  \vRp &  \eRp  \\
\enddata
\tablenotetext{a}{HAT--P--4 (Kovacs et al.\ 2007); HAT--P--6 (Noyes et al. 2008); \xon\ (this work).}
\tablenotetext{b}{Uncertainty is for \xon; those for HAT--P--4 and HAT--P--6 are similar or smaller.}
\tablenotetext{c}{Tabulated value has been rounded.}
\end{deluxetable}

\clearpage

\begin{thebibliography}


\bibitem[Bakos et al.(2007)]{2007ApJ...670..826B} Bakos, G.~{\'A}., et al.\ 
2007, \apj, 670, 826 

\bibitem[Barbieri et 
al.(2007)]{2007A&A...476L..13B} Barbieri, M., et al.\ 2007, \aap, 476, L13 


\bibitem[Burke et al.(2007)]{xo2b} Burke, Christopher J., et al.\ 2007, \apj, 671, 2115

\bibitem[Burke et al.(2008)]{xo5b} Burke, Christopher J., et al.\ 2008, \apj, submitted


\bibitem[Claret(2000)]{2000A&A...363.1081C} Claret, A.\ 2000, \aap, 363, 
1081 

\bibitem{Cochran(2000)} Cochran, W.\ 2000,
FTS spectrum of I2 Cell HRS3 at 69.9 C.,
ftp://nsokp.nso.edu/FTS\_cdrom/FTS50/001023R0.004

\bibitem[Cutri et al.(2003)]{twomass} Cutri, R.~M., et al.\ 
2003, The IRSA 2MASS All-Sky Point Source Catalog, NASA/IPAC Infrared 
Science Archive

\bibitem[Demarque et al.(2004)]{Dem04}
Demarque, P., Woo, J-H, Kim, Y-C, \& Yi, S. K. 2004, \apjs, 155, 667


\bibitem[Dravins et al. 1998]{drav98} Dravins, D., Lindegren, L., Mezey, E. \& Young, A. T. 1998, PASP, 110, 610


\bibitem[Fischer \& Valenti(2005)]{2005ApJ...622.1102F} Fischer, D.~A., \& 
Valenti, J.\ 2005, \apj, 622, 1102 

\bibitem[Gillon et 
al.(2007)]{2007A&A...472L..13G} Gillon, M., et al.\ 2007, \aap, 472, L13 


\bibitem[Harwit(1981)]{1981cdss.book.....H}
Harwit, M.\ 1981, Cosmic discovery. The search, scope, and heritage of astronomy,
Brighton: Harvester Press


\bibitem[H{\o}g et al.(2000)]{2000A&A...355L..27H} H{\o}g, E., et al.\ 2000, \aap, 355, L27 


\bibitem[Hinkle et al.(2000)]{echelle_reduction} Hinkle, K., Wallace, L., Valenti, J.,\& Harmer, D.\ 2000,
Visible and Near Infrared Atlas of the Arcturus Spectrum 3727-9300 A, ed.~Kenneth Hinkle,
Lloyd Wallace, Jeff Valenti, and Dianne Harmer.~(San Francisco: ASP) ISBN: 1-58381-037-4, 2000

 %

\bibitem[Johns-Krull et al.(2008)]{xo3b} Johns-Krull, C.~M., 
et al.\ 2008, \apj, 677, 657 


\bibitem[Kov{\'a}cs et al.(2007)]{hat4b} Kov{\'a}cs, G., et al.\ 2007, \apjl, 670, L41


 %
\bibitem[Mandel \& Agol(2002)]{2002ApJ...580L.171M} Mandel, K., \& Agol, 
E.\ 2002, \apjl, 580, L171 

\bibitem[Mandushev et al.(2005)]{2005ApJ...621.1061M} Mandushev, G., et 
al.\ 2005, \apj, 621, 1061 

\bibitem[McCullough, P. R., \& Burke, C. J.(2007)]{2007ASPC..366..70M}
McCullough, P. R. \& Burke, C. J.\ 2007, Astronomical Society of the Pacific Conference Series, 366, 70

\bibitem[McCullough et al.(2005)]{2005PASP..117..783M} McCullough, P.~R., 
Stys, J.~E., Valenti, J.~A., Fleming, S.~W., Janes, K.~A., \& Heasley, 
J.~N.\ 2005, \pasp, 117, 783 


\bibitem[Gould \& Morgan(2003)]{2003ApJ...585.1056G}
Gould, A., \& Morgan, C.~W.\ 2003, \apj, 585, 1056 

\bibitem[Morrison et al.(2001)]{gsc12}
Morrison, J.~E., R{\"o}ser, S., McLean, B., Bucciarelli, B., \& Lasker, B.\ 2001, \aj, 121, 1752 

\bibitem[Southworth(2008)]{2008MNRAS.tmp..444S} Southworth, J.\ 2008, \mnras, 444 


\bibitem{2005ApJ...619..558T}
Torres, G., Konacki, M., Sasselov, D.~D., \& Jha, S.\ 2005, \apj, 619, 558

\bibitem[Torres et al.(2008)]{2008ApJ...677.1324T} Torres, G., Winn, J.~N., \& Holman, M.~J.\ 2008, \apj, 677, 1324 


\bibitem[Tull(1998)]{1998SPIE.3355..387T} Tull, R.~G.\ 1998, \procspie, 
3355, 387 

\bibitem[Tull et al.(1995)]{1995PASP..107..251T} Tull, R.~G., MacQueen, 
P.~J., Sneden, C., \& Lambert, D.~L.\ 1995, \pasp, 107, 251 

\bibitem[Valenti \& Fischer(2005)]{2005ApJS..159..141V} Valenti, J.~A., \& 
Fischer, D.~A.\ 2005, \apjs, 159, 141 

\bibitem[Valenti \& Piskunov(1996)]{1996A&AS..118..595V} Valenti, J.~A., \& 
Piskunov, N.\ 1996, \aaps, 118, 595 


\bibitem[Wallace et al.(1998)]{solar_atlas}
Wallace, L., Hinkle, K., \& Livingston, W.\ 1998,
An atlas of the spectrum of the solar  photosphere from 13,500 to 28,000 cm-1 (3570 to 7405 \AA),
Publisher: Tucson, AZ: National Optical Astronomy Observatories

\bibitem[Winn et al.(2008)]{2008arXiv0804.4475W} Winn, J.~N., et al.\ 2008, 
ArXiv e-prints, 804, arXiv:0804.4475

 %

\end{thebibliography}
\end{document}